\input{aipcheck}

\documentclass[
    ,final            % use final for the camera ready runs
  ]
  {aipproc}

\layoutstyle{8x11single}
\usepackage{graphicx}
\usepackage{dcolumn}% Align table columns on decimal point
\usepackage{bm}% bold math
\usepackage{amssymb}
\usepackage{amsmath}
\usepackage{amsthm}
\usepackage{latexsym}

\newcommand{\be}{\begin{equation}}
\newcommand{\ee}{\end{equation}}
\newcommand{\bea}{\begin{eqnarray}}
\newcommand{\nn}{\nonumber}
\newcommand{\eea}{\end{eqnarray}}

\begin{document}

\title{Multipole moments as a tool to infer from gravitational waves the geometry around an axisymmetric body}


\author{Thomas P.~Sotiriou}{
  address={SISSA-International School of Advanced Studies, via Beirut 2-4, 34014, Trieste, Italy}, altaddress={INFN, Sezione di Trieste, Italy}
}
\author{Theocharis A.~Apostolatos}{address={National and Kapodistrian University of Athens, Panepistimiopolis, Zografos GR-15783, Athens, Greece}}
%\author{<author2>}{
%  address={<common address for author2 and author3>}
%}

\begin{abstract}
A binary system, composed of a compact object orbiting around a massive central body, 
will emit gravitational waves which will depend on the central body's spacetime geometry.
 We expect that the gravitational wave observables will somehow ``encode'' the information 
about the spacetime structure. On the other hand, it has been known for some time that the 
geometry around an axisymmetric body can be described by its (Geroch-Hansen) multipole moments.
 Therefore one can speculate that using the multipole moments can prove to be a helpful tool 
for extracting this information. We will try to demonstrate this in this talk, following the procedure 
described by [F.~D.~Ryan, Phys.~Rev.~D {\bf 52} 5707 (1995)] and 
[T.~P.~Sotiriou and T.~A.~Apostolatos, Phys.~Rev.~D {\bf 71} 044005 (2005)]. 
\end{abstract}

\maketitle

%%%%%%%%%%%%%%%%%%%%%%%%%%%%%%%%%%%%%%%%%%%%
%% MAINMATTER
%%%%%%%%%%%%%%%%%%%%%%%%%%%%%%%%%%%%%%%%%%%%

\section{\label{sec:1}Introduction}
%%%%%%%%%%%%%%%%%%%%%%
Gravitational waves  carry valuable information about the geometry of spacetime.
It is an interesting task to find ways to extract this information. At the same time we know that the 
geometry around a massive compact object is directly related to its Geroch-Hansen multipole moments \cite{Geroch,Hansen,Simon1,Simon2}.
Spacetime outside a stationary axisymmetric body is fully described by two sets of moments: the mass and
the mass-current moments. The first attempt to use the multipole moments as a tool to infer the geometry around a massive axisymmetric body through gravitational
waves was made by Ryan \cite{Ryan}. 

In \cite{sotapo} Ryan's results were generalized to include objects endowed with
strong electromagnetic fields.  In this case two additional families of moments are required to describe spacetime 
structure: the electric and the magnetic moments. It was shown that it is theoretically possible to use 
gravitational wave observables to infer all moments, and therefore get valuable information about the central
body's electromagnetic fields as well. 

For the purpose of demonstrating the above, a simplified but rather realistic model was used.
Consider a test particle orbiting a much more massive one. Such a system would emit gravitational radiation.
An example where such a model could be applied is a black hole binary with masses of the order of $10^6$ $M_\odot$ and $10^4$ $M_\odot$ respectively. 
Binaries with ratios of masses of this order will hopefully be observed by both Earth-based gravitational wave detectors 
and the forthcoming space detector LISA \cite{LIGOetal,LISA}.
The simplifying but rather realistic assumptions used here \cite{com} are the following:
Firstly, the central body is assumed to be stationary, axisymmetric and reflection symmetric with 
respect to the equatorial plane. It can, however, be endowed with an intense electromagnetic field. In this 
case it is characterized by four families of scalar multipole moments: its mass moments 
$M_l$, its mass-current moments $S_l$, its electrical moments $E_l$, and its magnetic moments 
$H_l$ where $l=0,1,2,\ldots$. Secondly, the orbiting object is 
assumed to be sufficiently small with respect to the central one so that it will move on a
 geodesic without distorting the geometry. Its orbit is assumed to be equatorial and circular
 or slightly deviating from that.
Thirdly, it is assumed that gravitational wave energy is given by the quadrupole formula since there
is no other way to make the required calculations analytically. Furthermore, this energy is assumed to be
carried away to infinity without any loss or absorption by the central object. 
 
Four measurable quantities  are presented as power series of the Newtonian orbital velocity of the
 test body with coefficients that are shown to be simple polynomials of the various moments. 
 Thus
 if any of these quantities are measured with sufficient accuracy, the lowest moments
 could be inferred and we could get valuable information 
about the internal structure of the compact massive body. The fact that the electromagnetic 
moments of spacetime can be measured demonstrates that one can obtain information about the 
electromagnetic field purely from gravitational wave analysis. Additionally, as will be discussed later, 
a possible deviation of the moments away from the Kerr-Newman moments corresponding to the same mass and 
angular momentum could reveal the presence of matter in the vicinity of the central object in the form of
a disk or a dust cloud.

Note at this point that many aspects of this work, some of which are taken for granted here, are discussed more thoroughly in \cite{sotapo} and 
the references therein.

\section{\label{sec:2}Observables}
%%%%%%%%%
%%%%%%%%%%%%%%%%%%%%%%
The quantities of interest, related to the motion of the test particle are the following: The energy per unit test-body mass for an equatorial circular orbit in an axially symmetric spacetime,
\be
\label{Eovermu}
\frac{E}{\mu}=\frac{ -g_{t t} - g_{t \phi} \Omega}
{\sqrt{-g_{t t}-2 g_{t \phi}\Omega - g_{\phi \phi} \Omega^2} } ,
\ee
and its orbital frequency,
\be
\label{Omega}
\Omega=\frac{ -g_{t \phi,\rho} +
\sqrt{(g_{t \phi,\rho})^2-(g_{tt,\rho})(g_{\phi \phi,\rho})} }{ g_{\phi \phi,\rho} }. 
\ee
The main frequency of the emitted gravitational waves, which is the frequency of the 
dominant spectral line, is given by the following formula:
\be
f=\frac{2 \Omega}{2\pi}=\frac{\Omega}{\pi}.
\ee
All of the above quantities can be expressed, as seen from these formulae, in terms of
the metric functions. We can relate these quantities to a number of observables.
The first one is the gravitational wave spectrum, i.e. the energy loss in gravitational waves per logarithmic interval of frequency.
\be
\label{delE}
\frac{\Delta E}{\mu}=-\Omega \frac{d (E/\mu)}{d \Omega}
\ee

If we assume that the orbit is not exactly equatorial but deviates slightly from that we expect to 
observe also two precession frequencies, the periastron precession frequency $\Omega_\rho$ and the orbital precession
frequency $\Omega_z$. They are given by
\bea
\label{omegas}
\Omega_\alpha = \Omega &-&
\left( -\frac{g^{\alpha\alpha}}{2} \left[
(g_{tt} + g_{t \phi} \Omega)^2
    \left(\frac{g_{\phi \phi}}{\rho^2} \right)_{,\alpha \alpha} \right. \right.\nn\\ &-&  2 (g_{tt} + g_{t \phi} \Omega)(g_{t \phi} + g_{\phi \phi} \Omega)
    \left(\frac{g_{t \phi}}{\rho^2} \right)_{,\alpha \alpha}% \nn \\
%&+ & 
+\left. \left. (g_{t \phi} + g_{\phi \phi} \Omega)^2
    \left(\frac{g_{t t}}{\rho^2} \right)_{,\alpha \alpha} {} \right] \right) ^{1/2}, 
\eea
when $\alpha$ takes the values $\rho$ and $z$ respectively.

Finally, the most interesting
observable is the number of cycles per logarithmic interval of frequency,\be
\label{delN}
\Delta N=\frac{f \Delta E(f)}{dE_{\textrm{wave}}/dt}.
\ee

 The reason for this is that it
is a quantity directly related to the phase evolution, which is the most accurately measured observable.
$dE_{\textrm{wave}}/dt$ is the gravitational wave luminosity, which is assumed to be given by the quadrupole
formula:
\be
\label{dEdtI}
\left. \frac{dE_{\textrm{wave}}}{dt} \right|_{I_{ij}}=\frac{32}{5} \mu^2 \rho^4 \Omega^6
\ee
plus a contribution from the current quadrupole radiative moment, due to the motion of the central object around the
center of mass (see Ref.~\cite{Ryan}),
\bea
\label{dEdtJ}
\left. \frac{dE_{\textrm{wave}}}{dt} \right|_{J_{ij}}&=&
\frac{32}{5} \left(\frac{\mu}{M}\right)^2 v^{10} %\times{}\nn\\
%& &{}\times
\left[ \frac{v^2}{36}-\frac{S_1 v^3}{12 M^2}+\frac{S_1^2 v^4}{16 M^4}+{\mathcal O}(v^5) \right],
\eea
where
\be
v\equiv (M\Omega)^{1/3}
\ee
is the Newtonian orbital velocity.
Some further additional contributions to $dE_{\textrm{wave}}/dt$,
due to post-Newtonian corrections should also be considered. The corresponding terms, up to $v^4$,
have coefficients that are simply numerical if one computes them from perturbation analysis in a Schwarzschild background.
In principle of course, for the symmetries of spacetime considered 
here, this perturbations analysis should be performed in a Kerr background.
One can compare though, the final formula (55) of \cite{Ryan} for $dE/dt$, and formula (3.13) of \cite{Shibetal}, which is
based on perturbations on a Kerr background. Such a comparison shows that at least up to order $v^4$, the terms of
\cite{Ryan} that include $S_1$ and $M_2$, which come from the corresponding contributions
that are given by Eqs.~(\ref{dEdtI},\ref{dEdtJ}),  in the case of a Kerr metric, are equal to the ones of Ref. \cite{Shibetal}.
This agreement indicates
that up to order $v^4$ we can simply add the following numerical post-Newtonian terms (the corresponding
terms of \cite{Shibetal} if we set $q=0$) onto the rest of the contributions to $dE_{\textrm{wave}}/dt$
\bea
\label{dEdtPN}
\left. \frac{dE_{\textrm{wave}}}{dt} \right|_{PN}&=&
\frac{32}{5} \left(\frac{\mu}{M}\right)^2 v^{10} %\times
\left[ -\frac{1247}{336}v^2 +{}\right.%\nn\\
%& &{}
+\left. 4\pi v^3 - \frac{44711}{9072} v^4 +{\mathcal O}(v^5) \right].
\eea
Finally, in order to compute the number of cycles $\Delta N$, one has to add up all of the contributions to
 $dE_{\textrm{wave}}/dt$ and combine them with the expression for $\Delta E/\mu$,
 which is given above (see Eq.~(\ref{delE})).

\section{\label{sec:3} Multipole moments}
%%%%%%%%%
%%%%%%%%%%%%%%%%%%%%%%

As already mentioned we are focusing on stationary axisymmetric spacetimes. The most general metric 
for those symmetries is the Papapetrou metric \cite{pap},
\be
\label{papmet}
ds^2=-F(dt-\omega ~d\phi)^2+\frac{1}{F} \left[
e^{2\gamma} (d\rho^2 + dz^2) + \rho^2 d\phi^2 \right],
\ee
where $F$, $\omega$ and $\gamma$ are functions of just $\rho$ and $z$.

All of the observables mentioned can be expressed in terms of the metric functions alone. On the other hand,
we know that the metric can be expressed in terms of the multipole moments. Ernst has shown long ago
that the metric functions are fully determined by two complex functions, the well known Ernst potentials $\mathcal{E}$ and $\Phi$ \cite{ernst1,ernst2}, where
\bea
\label{ernstpot}
\mathcal{E}&=&(F-\left|\Phi\right|^2)+i\varphi,
%=\frac{\sqrt{\rho^{2}+z^{2}}-\tilde{\xi}}{\sqrt{\rho^{2}+z^{2}}+\tilde{\xi}}\nn\\
%\Phi&=&\frac{\tilde{q}}{\sqrt{\rho^{2}+z^{2}}+\tilde{\xi}}
\eea

\bea
\label{intgtf}
g_{t\phi}=F \omega&=& F \int_{\rho}^{\infty} \!\!d\rho'
\frac{\rho'}{F^{2}}\left(\frac{\partial\varphi}{\partial z}+
2{\Re}(\Phi)\frac{\partial{\Im}(\Phi)}{\partial z}-{}\right.%\nn\\
%& &{}
-\left.\left.
2{\Im}(\Phi)\frac{\partial{\Re}(\Phi)}{\partial z}
\right)\right|_{z=\textrm{const}}.
\eea
Instead of these one can also use the functions $\tilde{\xi}$ and $\tilde{q}$, defined by the relations
\bea
\label{etoxi}
\mathcal{E} = \frac{\sqrt{\rho^{2}+z^{2}}-\tilde{\xi}}{\sqrt{\rho^{2}+z^{2}}+\tilde{\xi}},\quad
\Phi=\frac{\tilde{q}}{\sqrt{\rho^{2}+z^{2}}+\tilde{\xi}},
\eea 
which are more directly related to the moments.

These functions have two very interesting attributes. Firstly, they can be 
written as power series expansions of 
$\rho$ and $z$ at infinity 
\bea
\label{expxiq}
\tilde {\xi } = \sum\limits_{i,j = 0}^\infty {a_{ij}\bar {\rho }^i\bar {z}^j},\qquad
\tilde {q} = \sum\limits_{i,j = 0}^\infty {b_{ij}\bar {\rho }^i\bar {z}^j}
\eea
where
\be
\label{barred}
\bar \rho \equiv \frac{\rho}{\rho^2+z^2},\qquad \bar z \equiv \frac{z}{\rho^2+z^2}.
\ee
Secondly, they can be fully determined by their values on the symmetry
axis:
\bea
\tilde{\xi}(\bar\rho=0)=\sum_{i=0}^{\infty} m_i {\bar z}^i \qquad
\tilde{q}(\bar\rho=0)=\sum_{i=0}^{\infty} q_i {\bar z}^i.
\eea

The coefficients $m_i$ and $q_i$ are related to the mass moments, $M_i$, 
the mass current moments, $S_i$, the electric field moments, $E_i$, 
and the magnetic field moments, $H_i$ through the recursive algebraic relations:
\bea
\label{lom}
m_{n}&=&a_{0n}=M_{n}+iS_{n}+\textrm{LOM}\nn\\
q_{n}&=&b_{0n}=E_{n}+iH_{n}+\textrm{LOM}
\eea
The term LOM is an abbreviation for ``Lower Order Multipoles''. The full version of eqs.~(\ref{lom}) and their derivation can be
found in \cite{SotiApos,sotapo}.

\section{\label{sec:4}Algorithm and expansion folmulae}
%%%%%%%
%%%%%%%%%%%%%%%%%%%%
Having expressed the observables in terms of the metric functions and the metric functions in terms
of the moments, what is left to do is a mathematical manipulation, which, however, is not straightforward.
The algorithm one follows to express the observables in terms of the moments is the following:
\begin{enumerate}
\item Use the equations of the previous section and express the metric functions 
as power series in $\rho$ and $z$, with coefficients depending only on the multipole moments.
\item The observable quantities discussed depend only on the metric functions. Thus they can also be expressed 
as power series in $\rho$, with coefficients depending only on the multipole moments; $z$ is no longer
present in the expansions since the evaluation takes place on the equatorial plane.
\item Use the assumed symmetry to simplify the computation. $M_i$ is zero when $i$ is odd and $S_i$ is 
zero when $i$ is even. For electromagnetic moments there are two distinct cases: $E_i$ is zero for odd $i$ and $H_i$ is zero for even $i$ or vice versa.
 We will call the first case electric-symmetric (es) since the electric field respects reflection symmetry 
and the second case magnetic-symmetric (ms) for the equivalent reason (see Section IIC of \cite{sotapo} for a detailed 
analysis of the symmetries of the electromagnetic fields and their effect on the moments).
\item Finally, since $\Omega=\Omega(\rho))$ we can invert and express $\rho$, and consequently the observables, as a series in $\Omega$, or
$v\equiv (M\Omega)^{1/3}$.
\end{enumerate}

Following this algorithm we have expressed all of the quantities as power series in $v$ with coefficients
depending only on the moments: 

\be
\label{wpprosw}
\frac{\Omega_{\rho}}{\Omega}=\sum\limits_{n=2}^{\infty}R_{n}v^{n},\qquad \frac{\Omega_{z}}{\Omega}=\sum\limits_{n=3}^{\infty}Z_{n}v^{n},\qquad
%\ee
%\be
%\label{deprosm} 
\frac{\Delta E}{\mu}=\sum\limits_{n=2}^{\infty}A_{n}v^{n}
\ee
Due to lack of space, the full expressions for $R_n$ and $Z_n$ are not given here. They can be found, however, in \cite{sotapo}.
The first coefficients of the energy per unit mass, $\Delta E/\mu$, have been chosen as a representative example .
The results are similar  for $R_n$ and $Z_n$ both in the electric-symmetric
 case (es) and in the magnetic-symmetric case (ms). The first seven coefficients are:
\bea
\label{deprosm1}
A^{(es)}_{2}&=&\frac{1}{3},\quad
A^{(es)}_{3}=0,\quad
A^{(es)}_{4}=-\left(\frac{1}{2}-\frac{2}{9}\frac{E^{2}}{M^{2}}\right),\nn\\
A^{(es)}_{5}&=&\frac{20}{9}\frac{S_{1}}{M^{2}},\quad
A^{(es)}_{6}=-\frac{27}{8}+\frac{M_{2}}{M^{3}}+\frac{3}{2}\frac{E^{2}}{M^{2}}+\frac{1}{3}\frac{E^{4}}{M^{4}}\nn\\
A^{(es)}_{7}&=&\frac{28}{3}\frac{S_{1}}{M^{2}}-\frac{28}{9}\frac{H_{1}E}{M^{3}}+
\frac{56}{27}\frac{S_{1}E^{2}}{M^{4}}
\eea

Now, as already mentioned, the most accurately measured, and 
hence the most interesting, observable is the number of cycles per logarithmic interval of frequency, 
\be
\label{powserN}
\Delta N=\frac{5}{96 \pi} \left( \frac{M}{\mu} \right) v^{-5}
\left( 1 + \sum\limits_{n=2}^{\infty} N_{n} v^{n} \right).
\ee
The coefficients $N_n$ are given explicitly up to fourth order; all higher order coefficients
are given by recursive relations:
\bea
\label{Nfinal}
%N^{(es)}_1&=&0\nn\\
N^{(es)}_2&=&\frac{743}{336}+\frac{14}{3}\frac{E^{2}}{M^{2}},\quad\\
N^{(es)}_3&=&-4\pi+\frac{113}{12}\frac{S_{1}}{M^{2}},\quad\\
N^{(es)}_4&=&\frac{3058673}{1016064}-\frac{1}{16}\frac{S_{1}^{2}}{M^{4}}
+ 5\frac{M_{2}}{M^{3}}+  \frac{12431}{504}\frac{E^{2}}{M^{2}}
+\frac{179}{9}\frac{E^{4}}{M^{4}},\\
N^{(es)}_{4k+1}&=&
(-1)^{k}\frac{(16k+20)}{3}\frac{(2k-1)!!}{(2k-2)!!}
\frac{H_{2k-1}E}{M^{2k+1}}  +\textrm{LOM},\quad\\
N^{(es)}_{4k+2}&=&
(-1)^{k}\frac{(8k+6)(2k+2)}{3}\frac{(2k-1)!!}{(2k)!!}
\frac{E_{2k}E}{M^{2k+2}}    +\textrm{LOM}\\
N^{(es)}_{4k+3}&=&
(-1)^{k}\frac{(16k+28)}{3}\frac{(2k+1)!!}{(2k)!!}
\frac{S_{2k+1}}{M^{2k+2}}    +\textrm{LOM},\quad\\
N^{(es)}_{4k+4}&=&
(-1)^{k}\frac{(8k+10)}{3}\frac{(2k+3)!!}{(2k+2)!!}
\frac{ M_{2k+2} }{ M^{2k+3} } +\textrm{LOM}
\eea
Notice that a pattern is revealed for the appearance of the different moments in 
different order coefficients.
As we go to higher orders, higher order moments are present, but each time only one new moment appears that was
not present in the lower order coefficients. Through the recursive relations we can see that this holds for all
orders. The same is true for the other observables and it shows that the multipole moments can be extracted from accurate
measurements of these observables.

\section{\label{sec:5}Conclusions}
%%%%%%%%%%%%%%%%%%%%
%%%%%%%%%%%%%%%%%%%%%%%%%%%%%%%%%
As demonstrated above, it is possible to evaluate the moments from certain observables. Of course, the
accuracy with which observed quantities are measured affects the accuracy of the moment 
evaluation. As Ryan showed, LIGO is not expected to provide sufficient accuracy 
but LISA is very promising and we expect to be able to measure a few lower moments \cite{Ryan2}.These lower order moments can give valuable information: 
we can infer the mass, the angular momentum, the magnetic dipole, the overall charge, 
the quadrupole moment etc.  For 
a more thorough discussion of extracting the moments from measurements of the observables see Section IV
of \cite{sotapo}.

A crucial test that could be made once the moments have been evaluated is to check whether they are 
interrelated as in a Kerr-Newman metric \cite{SotiApos}. A negative outcome could lead to a number of different
conclusions. The most important are the following: the
 central object might not be a black hole but a neutron star.  However,
if its mass is high enough to exclude this possibility then 
the safest assumption is that the object is indeed a black hole but matter is present
 in its vicinity (e.g. an accretion disk). Studying how much and in which way the moments are deviating from the Kerr-Newman
ones, some first predictions could be made about the shape (e.g. disk, dust cloud, etc.) and the characteristics 
 (e.g. magnetic field) of this matter.

\begin{theacknowledgments}
The authors would like to thank John Miller for a critical reading of this manuscript and for constructive suggestions. The research presented here was supported in part by Grant No 70/4/4056 of the special Account for Research Grants of the University of Athens, and in part by Grant No 70/3/7396 of the ``PYTHAGORAS'' research funding program.
\end{theacknowledgments}



\bibliographystyle{aipproc}   % if natbib is available

%\bibliography{sample}

\begin{thebibliography}{20}
\bibitem{com} See Section I of \cite{sotapo} and references therein for a discussion on how realistic
these assumptions are.
\end{thebibliography}

%  \typeout{******************************************}
%  \typeout{** Please run "bibtex \jobname" to optain}
%  \typeout{** the bibliography and then re-run LaTeX}
%  \typeout{** twice to fix the references!}
%  \typeout{******************************************}

\end{document}